\documentclass[aps,prper,twocolumn,superscriptaddress,letterpaper,longbibliography]{revtex4-1}

\usepackage{graphicx}
\usepackage{mdframed}
\usepackage{framed}
\usepackage{indentfirst}
\usepackage{natbib}
\usepackage{amsmath,amssymb}
\usepackage{subfigure}
\usepackage{enumitem}
\usepackage{multirow}
\usepackage{epstopdf}

\begin{document}

\title{Retention of conceptual learning after an interactive introductory physics course}

\author{Bethany R. Wilcox}
\affiliation{Department of Physics, University of Colorado, 390 UCB, Boulder, CO 80309}

\author{Steven J. Pollock}
\affiliation{Department of Physics, University of Colorado, 390 UCB, Boulder, CO 80309}

\author{Daniel R. Bolton}
\affiliation{Department of Physics, University of Colorado, 390 UCB, Boulder, CO 80309}

\begin{abstract}
The cyclic format of the undergraduate physics curriculum depends on students' ability to recall and utilize material covered in prior courses in order to reliably build on that knowledge in later courses.  However, there is evidence to suggest that people often do not retain all, or even most, of what they learned previously.  How much information is retained appears to be dependent both on the individuals' approach to learning as well as the style of instruction.  In particular, there is evidence to suggest that active engagement techniques in the classroom can improve students' retention of the material over time.  Here, we report the findings of a longitudinal investigation of students' retention of conceptual understanding as measured by the Force and Motion Conceptual Evaluation (FMCE) following a first-semester, calculus-based introductory physics course, which features significant active engagement in both lecture and recitation.  By administering the FMCE at the end of a first-semester physics course and again at the beginning of the subsequent second-semester physics course, we examine students' knowledge retention over time periods ranging from 1-15 months.  We find that the shift in students' FMCE scores between these two courses is positive but corresponds to a small effect size, indicating that students retained effectively all of their conceptual learning (as measured by the FMCE).  This finding largely persists even as the length of the gap between the two courses increases.  We also find that, when breaking out students' performance on individual questions, the majority of students maintain their score on individual questions.  Averaged over all questions, roughly a fifth of the students switched their answers from right to wrong or wrong to right on any given item.  
\end{abstract}

\maketitle

\section{\label{sec:intro}Introduction and Motivation}

The undergraduate physics curriculum has historically featured a cyclic progression in which students return to core topical areas multiple times over the course of the program.  Each course targeting a particular topical area (e.g., classical mechanics) builds off what was taught in previous courses with the goal of digging deeper into the content, typically with increased conceptual and mathematical rigor.  To work, this process depends on students retaining a significant fraction of the material they learned in prior courses.  Unfortunately, there is significant evidence from educational psychology literature to suggest that students often do not retain much of what they learn over time \cite{bacon2006retention, rubin1996forgetting}.  A lack of retention often forces instructors to dedicate significant class time to catching students up on material they have seen in previous courses, undercutting the cyclic process of building on students knowledge.  

Literature on student retention also suggests that how much information a person retains depends on the nature of the information.  For example, memorized facts show low retention that drops off sharply over time \cite{bacon2006retention, conway1992retention}.  Alternatively, conceptual learning, which is reflective of deeper understanding, appears to be more robust and falls off more slowly over time \cite{bacon2006retention}.  Given this, it might be expected that courses designed to promote deeper conceptual understanding would result in greater retention of the material between courses.  Active engagement techniques, in particular, have produced well documented improvements in students conceptual understanding \cite{freeman2014active}, and, in some cases, have also been shown to improve students' retention of the material over time \cite{francis1998retention, bernhard2000ieretention, deslauriers2011retention}.  

At the University of Colorado Boulder (CU), the introductory physics sequence has been modified to include significant active engagement, including interactive lectures and fully interactive recitation sections \cite{pollock2005tutorials}.  Existing literature suggests that such a course, which is designed to encourage deep conceptual learning, may also result in high rates of retention.  Building on this literature, this longitudinal study investigates the following research questions:

\begin{itemize}
\item[1)]  How much conceptual knowledge -- as measured by the Force and Motion Conceptual Evaluation (FMCE) \cite{thornton1998fmce} -- do students retain through the gap between two of the courses in our introductory physics sequence?
\item[2)]  Does the level of retention correlate with the time duration of the gap between the two courses? 
\item[3)]  Does the level of retention vary significantly between different items on the FMCE?
\end{itemize}

Here, we address the above research questions using two semesters of longitudinal data from students in CU's introductory physics sequence.  We begin by reviewing the existing literature on knowledge retention across multiple fields including physics education research (PER) (Sec.~\ref{sec:background}).  We then discuss the context and methods of the current study (Sec.~\ref{sec:context}).  Finally, we present results (Sec.~\ref{sec:results}), and discuss limitations and future work (Sec.~\ref{sec:discussion}).

\section{\label{sec:background}Background}

Issues around memory and knowledge retention have been studied previously both within the PER community, as well as in the broader education and psychology literature.  Many of the educational psychology studies show a sharp drop off in knowledge immediately following the original learning followed by a slower tail of knowledge loss over time, and much of this work focuses on quantifying the shape of this retention curve \cite{wixted1991form,rubin1996forgetting,bacon2006retention}.  However, these studies also show that retention over long intervals (intervals of a few weeks or longer) can vary significantly based on the nature of the information being recalled \cite{rubin1996forgetting,conway1992retention}.  For example, Bacon and Stewart \cite{bacon2006retention} examined business students' retention of information learned in a consumer behavior course.  They examine students performance on a post-test assessment at the end of the course and then again in a followup course taken later in the program.  The post-test was designed to include items targeting both rote memorization and deeper learning that required application of concepts or problem solving.  They found that students retained more of the knowledge targeted by the deep learning questions than that targeted by the rote memorization questions.  

In addition to showing differences in knowledge retention based on the type of knowledge being recalled, there is also evidence to suggest that the level of knowledge retention can be impacted by how the material was originally taught \cite{bjork1994memory}.  For example, Jackson \emph{et al.} \cite{jackson2004tutor} investigated the impact of an online AutoTutor program on students' learning and retention of introductory physics content.  They found that students who engaged with the AutoTutor program displayed greater retention of the material one week later than students who engaged only with an automated information delivery system that did not feature the same interactive tutoring features.  Similarly, Johnson \emph{et al.} \cite{johnson2016embodied} investigated the impact of three digital platforms, each of which encouraged different levels of embodied learning through sensorimoter experiences.  They found an interaction in which students in conditions featuring high levels of embodiment demonstrated greater retention of knowledge a week after the initial intervention.  

Literature from the PER community has also shown a high degree of long-term knowledge retention of students conceptual knowledge.  For example, Deslauriers and Wieman \cite{deslauriers2011retention} compared students retention of quantum mechanics conceptual knowledge after a modern physics course taught in a traditional lecture format and another taught in an interactive format.  They found that, while the initial gains in conceptual knowledge were larger for the students in the interactive class, students in both conditions showed almost complete retention of that knowledge in a followup course taken 6-18 months later.  The results of this study were also consistent with other studies, which showed almost no decline in students introductory physics knowledge as measured by an introductory conceptual assessment over periods of 1-3 years when these students experienced interactive introductory physics courses that utilized the \emph{Tutorials in introductory physics} curriculum \cite{mcdermott2002tutorials,pollock2009longitudinal,francis1998retention}.  Other studies have shown larger drops in students' retention over time, even when exposed to interactive curricula designed to promote conceptual learning.  For example, researchers at the University of Washington have observed differing results in terms of students' retention of knowledge around circuit behavior, with one study showing a roughly 25\% drop in students' performance in a longer-term followup assessment \cite{shaffer1992circuits}, while a followup study showed no drop in students' performance \cite{mcdermott2000circuits}.  

In this work, we contribute to the body of literature described above.  Specifically, we examining students' retention of their conceptual knowledge of introductory mechanics content over periods ranging from 1-15 months.

\section{\label{sec:context} Context \& Methods}

All data for this study were collected in an introductory mechanics course at University of Colorado Boulder (PHYS I), and the subsequent introductory electricity and magnetism course (PHYS II).  Both PHYS I and II are calculus-based course that feature interactive lectures which consistently utilize clicker questions \cite{crouch2001pi}, as well as fully interactive recitation sections that utilized the \emph{Tutorials in Introductory Physics} curriculum \cite{mcdermott2002tutorials}.  Both of these courses also utilize the online homework platforms \emph{Mastering Physics} \cite{MasteringPhysicswebsite} and \emph{FlipIt Physics} \cite{flipitwebsite}.  The population of these courses are primarily engineering and other science majors along with physics majors, though the physics majors are the minority.  The courses enroll anywhere from 800-1200 students per semester.  While we do not have demographic information on students in our sample, the demographics of these courses reflect both the demographics of CU, which is a predominantly white institution, and the demographics of physics generally, which typically has an over-representation of men.  

We utilized the Force and Motion Conceptual Evaluation (FMCE) \cite{thornton1998fmce} to measure students' conceptual learning associated with PHYS I.  The FMCE has been given pre- and post-instruction in PHYS I as a standard part of the course for more than a decade, and gains realized by this course are well above the national averages for courses taught in a traditional lecture format \cite{pollock2005tutorials,vonkorff2016pedagogy}.  For students in PHYS II, we also have post-instruction scores on the Brief Electricity and Magnetism Assessment (BEMA) \cite{ding2006bema}, which is given each semester as a standard part of PHYS II.  Faculty also rotate through courses, so with a few exceptions the instructors for any given semester of PHYS I and II were not the same.  

To collect information about students' retention of material from PHYS I, we also gave the FMCE as a pre-test in PHYS II during two semesters, one fall and one spring. The two semesters of FMCE data collection in PHYS II took place three semesters apart, allowing sufficient time for the majority of students from the most recent PHYS I semester to complete PHYS II even if they did not go directly from PHYS I to PHYS II.  As these two courses are the first two in a sequence and prerequisites for nearly ever other physics course within the CU physics and engineering programs, these students should not have had other direct exposures to the material on the FMCE in the intervening time between the two courses.  The only exception to this would be studying done by the student to prepare for the final in PHYS I which takes place approximately 1 week after the FMCE post-test is administered.  While this study will focus on the students in PHYS II as the primary population, it is worth noting that there is a significant selection effect at play with respect to which students persist from PHYS I to PHYS II, as evidenced by a roughly 6\% lower average score among the unmatched PHYS I students relative to those that have matched scores in PHYS II.

After eliminating responses that were invalid (less that 1-2\% of total received responses), we collected $N=1445$ responses to the FMCE in PHYS II.  Responses were marked as invalid if they left 6 or more questions blank.  To match students' FMCE scores between PHYS I and II, we pulled from the three semesters of PHYS I prior to the PHYS II semester in which we collected FMCE pre-data.  In this way, students who took a 1-2 semester gap between PHYS I and II could still be captured in our matched dataset.  Students who took PHYS I multiple times were matched to their most recent FMCE score.  Ultimately, the matched dataset captured 74\% ($N=1068$) of the full set of PHYS II students for whom we had valid FMCE responses.  The unmatched students include students who did not complete the FMCE post-test in PHYS I as well as students who did not take PHYS I at CU within the three semesters prior to taking PHYS II.  Table \ref{tab:gaps} shows the distribution of students across the various gap sizes in our dataset.

\begin{table}
\caption{Approximate length, in months, of the gap in time between taking the FMCE post-test in PHYS I and the FMCE pre-test in PHYS II, along with the number of students in the dataset with that gap duration.  }\label{tab:gaps}
 \begin{ruledtabular}
    \begin{tabular} {l c c c c c}
		Gap length (Months): & 1&3&8&13&15 \\
		\hline
		$N_{matched}$ &	532 & 343 & 130 & 26 & 37 	\\
	\end{tabular}
 \end{ruledtabular}
\end{table}

To determine what, if any, systematic differences might exist between students in the matched and unmatched data sets, we compare their performance on several other measures including the BEMA post-test and their final grade in PHYS II.  Students in the un-matched dataset scored lower, on average, on the FMCE in PHYS II than students in the matched dataset by 13.5\%, and this difference was statistically significant (Mann-Whitney U test, $p<0.001$).  This difference suggests a mismatch in the incoming preparation of the PHYS II students for whom we did not have a matched PHYS I FMCE score.  Students in the unmatched dataset also scored 0.7\% higher on the BEMA, which was administered at the end of PHYS II, and 3.6\% lower in overall average final course score; only the difference in overall average course score was statistically significant (Mann-Whitney U test, $p<0.05$) and of moderate size ($d=0.3$).  This suggests that the unmatched students appeared to have been less prepared with respect to the mechanics content tested by the FMCE and performed, on average somewhat lower in terms of overall course score; however, the differential in overall course score was considerably smaller than that on the FMCE. 

To better understand the makeup of the students in the unmatched dataset, we examined rosters for all of the semesters of PHYS I included in the study to determine if the unmatched students took PHYS I at CU and simply missed the FMCE post-test for some reason.  Of the 387 unmatched PHYS II students, 181 of them appeared in a PHYS I roster from a previous semester.  The remaining 206 students are primarily students who did not take PHYS I at CU -- for example, transfer students or students who came into college with physics credit from high school.  Comparing the performance on the FMCE in PHYS II between the unmatched students who took PHYS I at CU and those who did not, we saw slightly lower performance in the group who did not take PHYS 1 at CU (an average score of 63.8\% compared to 67.2\%), but this difference was not statistically significant (Mann-Whitney U test, $p=0.2$), though the smaller $N$ in the unmatched dataset also limits our statistical power.  This suggests that, while students who take PHYS I at CU may realize some small benefit in terms of FMCE performance over those who took PHYS I elsewhere, this benefit is not the only, or even the primary, factor contributing to the difference in FMCE scores observed in our matched and unmatched datasets.

\section{\label{sec:results}Results}

In this section, we present findings organized according to each of the three major research questions outlined in Sec.~\ref{sec:intro}.  

\begin{figure*}
\includegraphics[width=\linewidth]{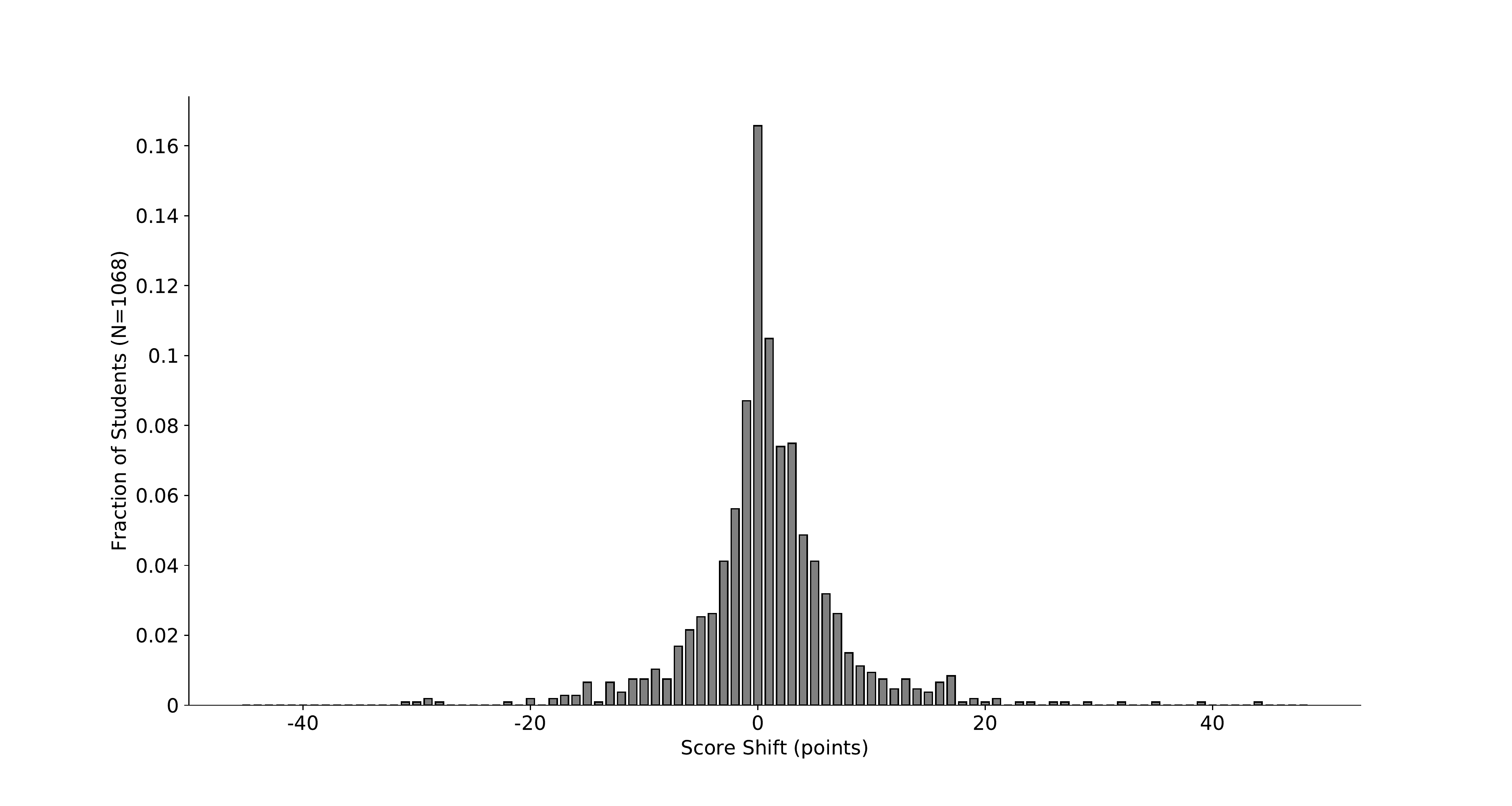}
\caption{Distribution of the shift in students' raw FMCE scores between PHYS I and PHYS II.  Bins are 1 point in size and the bin is centered on the value of the shift.  Positive indicates the student's score was higher in PHYS II than in PHYS I.  Note the distribution is not normal due to the large peak at a score shift of zero (Skewness-Kurtosis test, $p<0.001$).  There are 47 questions on the FMCE resulting in a maximum possible shift of 47 points in either direction.  }\label{fig:histogram1}
\end{figure*}

\begin{figure*}
\includegraphics[width=\linewidth]{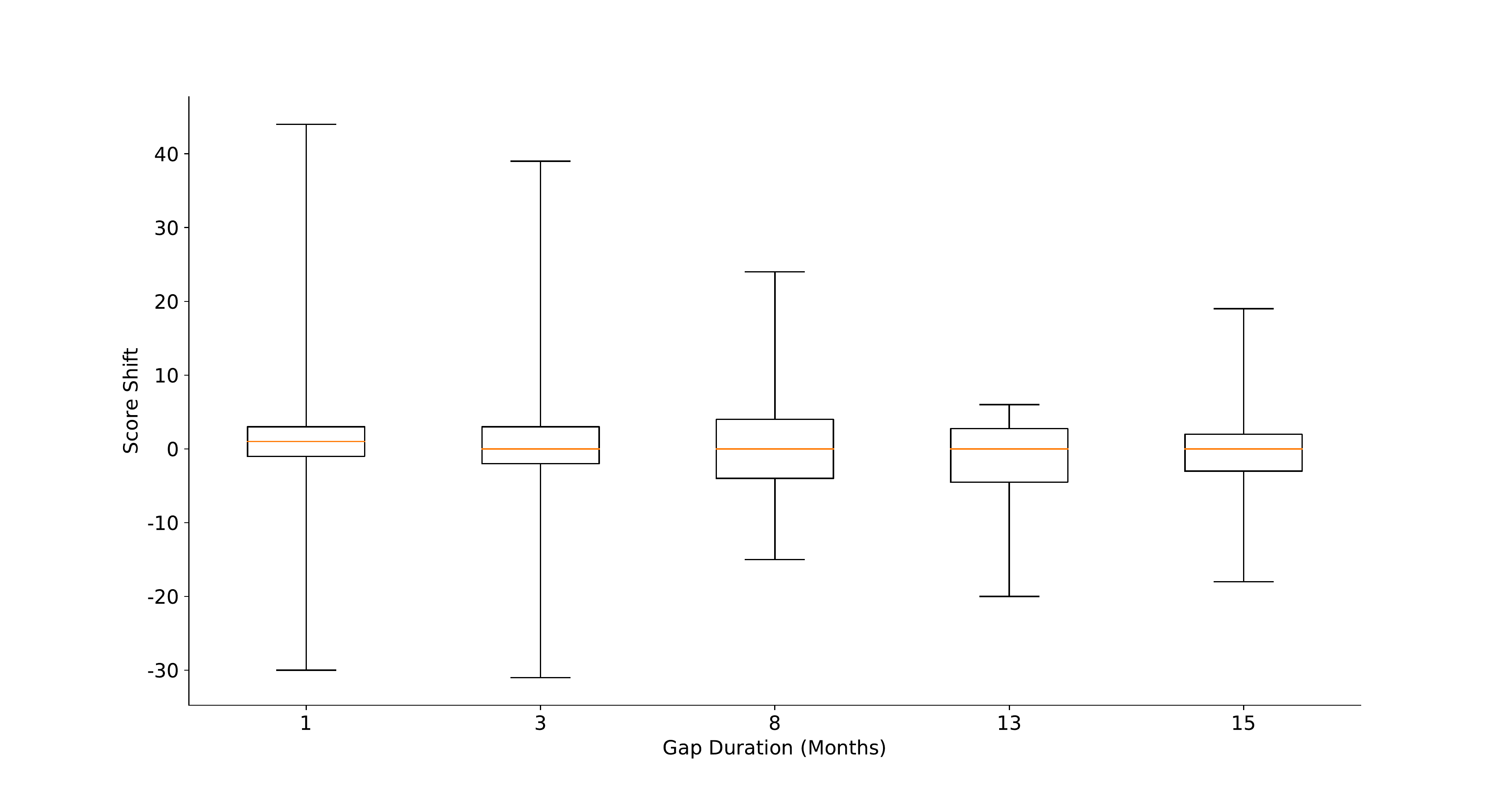}
\caption{Box and whisker plots of the FMCE raw score shift for each of the five gap duration in the dataset.  The center line for each boxplot represents the median shift, while the box denotes the interquartile range (meaning 50\% of responses fall within this range) and the whiskers show the max and min shift value (with a max possible shift of 47 points in either direction). }\label{fig:boxplots}
\end{figure*}

\begin{figure*}
\includegraphics[width=\linewidth]{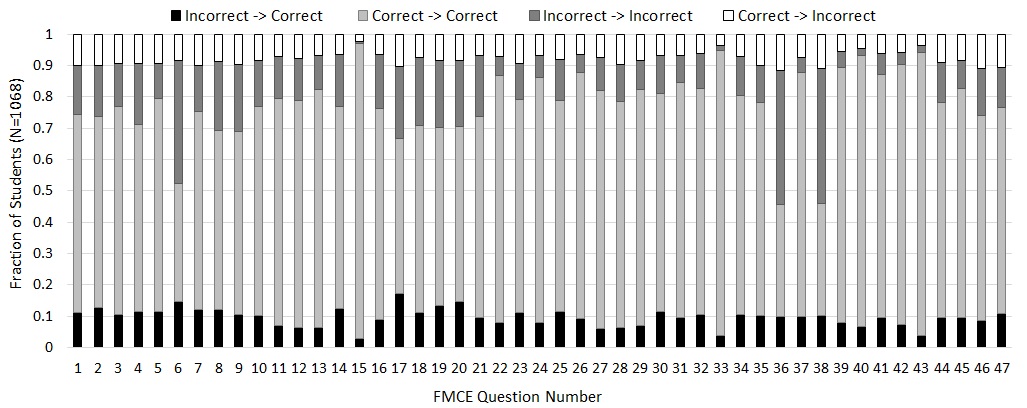}
\caption{Fraction of students who exhibited any of the following four response patterns in their responses to each FMCE item between PHYS I and II: 1) an incorrect response in PHYS I that shifted to a correct response in PHYS II, 2) an incorrect response in PHYS I that stayed an incorrect response in PHYS II, 3) a correct response in PHYS I that stayed a correct response in PHYS II, or 4) a correct response in PHYS I that shifted to an incorrect response in PHYS II. }\label{fig:itemShifts}
\end{figure*}

\subsection{\label{sec:overall}Overall retention of mechanics knowledge on the FMCE}

Here, we report the FMCE score as the sum of items scores over all 47 questions.  When the analysis is run using the scoring proposed by Thornton \cite{thornton2009fmce}, the trends of the analysis do not change.  Figure \ref{fig:histogram1} shows a histogram of the shift in students' FMCE scores between PHYS I and PHYS II.  This histogram is not normally distributed (Skewness-Kurtosis test, $p<0.001$) primarily because of the large spike in the number of students with an overall shift of zero.  The average shift was 0.87 points (with a maximum possible shift magnitude of 47 points), meaning the students got roughly one more question right on the FMCE in PHYS II than they did in PHYS I.  The size of the matched data set provides significant statistical power, and the difference between the PHYS II and PHYS I score distributions was statistically significant (Mann-Whitney U test, $p<0.001$); however, the difference in means represented an effect size of only $d=0.1$, which represents a small effect size with negligible practical significance.  This slight increase in score may be attributable to the impact of students studying for the final exam in PHYS I in the week between when the FMCE and final exam take place.  This finding is consistent with previous literature around students' retention of conceptual knowledge after going through an interactive introductory physics course \cite{pollock2009longitudinal, francis1998retention}.

\subsection{Impact of gap duration on conceptual knowledge retention}

As shown in Table \ref{tab:gaps}, students in the matched PHYS II data set experienced gap lengths of between 1 and 15 months.  To determine if the length of the gap is related to students' retention on the FMCE, we correlated the shift in their FMCE score with the duration of the gap.  We found a Spearman correlation coefficient of $r=-0.1$ that, while quite small, was statistically significant ($p<0.001$).  To lend more insight into this correlation, Fig.~\ref{fig:boxplots} also shows a comparison of students performance for each of the five gap durations included in this study.   Figure \ref{fig:boxplots} shows that there is very little variation in the median shifts as the gap size increases, though the overall range in the scores tends to decrease after the first 1-3 semesters.  

\subsection{Retention by item on the FMCE}

There are at least two possible explanations for the lack of significant shift in the overall FMCE score between PHYS I and II.  One possibility is that basically all students answer the majority of the FMCE questions the same way both times, while another option is that there is significant variation in students individual responses that cancels out on average.  To distinguish between these two possibilities, we examined students responses by item to determine if their score on that item shifted between PHYS I and II.  Figure \ref{fig:itemShifts} shows the fraction of students whose responses fell into each of the following four possible response patterns: 1) an incorrect response in PHYS I that shifted to a correct response in PHYS II, 2) a correct response in PHYS I that stayed a correct response in PHYS II, 3) an incorrect response in PHYS I that stayed an incorrect response in PHYS II, or 4) a correct response in PHYS I that shifted to an incorrect response in PHYS II.  

Averaged across all items, 82.4\% of students gave consistent responses to the item in PHYS I and II by responding correct for both or incorrect for both.  Recall that both of these test are post-instruction and students had no additional instruction in the material between taking the FMCE in PHYS I and re-taking it in PHYS II.  Figure \ref{fig:itemShifts} also shows that there is relatively little variation in the level of consistency across all the questions on the FMCE; the question with the highest consistency of responses was Q15 (94.7\% consistent), followed by Q33 (92.4\% consistent), while the question with the lowest consistency was Q17 (72.5\% consistent).  Q15 and Q33 on the FMCE are the two easiest questions that typically demonstrate $\sim$95\% correct before students take PHYS I; as such, we have historically used these questions used as a gauge to judge how seriously students are taking the exam.  Q17 asks about the force on an object moving at a constant velocity and represents one of the more difficult questions on the exam.

\section{\label{sec:discussion}Discussion \& Limitations}

Here, we examined students retention of conceptual mechanics knowledge, as measured by the FMCE, between a first- and a second-semester physics course at a large, research university.  Both of these courses were highly interactive and designed to maximize students' conceptual learning.  We found that on average students' FMCE scores increased by just under one point between the two courses and that this result did not vary significantly based on the length of the time that elapsed between the two courses.  There are several factors that might explain the slight increase in students' scores between PHYS I and PHYS II.  First, students spend approximately one week studying for the PHYS I final after the FMCE is given.  Second, students are likely less stressed and burnt out at the beginning of the semester than at the end of the semester just before finals week.  And finally, a small subset of the PHYS II students in our dataset may have been taking PHYS II for the second time after failing to complete or pass it in a previous semester.  All of the above factors could contribute to a slight increase in the average performance in PHYS II over PHYS I despite the matched nature of the dataset.  We also found that, on average, just over 80\% of students gave consistent answers to each individual item when taking the FMCE the second time.  Taken together, these results suggest that the conceptual learning gained in PHYS I was almost entirely retained over the gap between the courses.  This result is consistent with other results in the literature that suggests that conceptual learning is robust and sustained over time \cite{pollock2009longitudinal, francis1998retention}.  

The main findings described above were based on those students for whom we had matched post-FMCE scores in PHYS I and pre-FMCE scores in PHYS II.  This represented roughly three quarters of the students for whom we had pre-FMCE scores in PHYS II.  Comparisons of student performance between the matched and unmatched students in our data set showed that the unmatched students scored close to 15\% lower on the FMCE but had no significant difference in their performance on the BEMA at the end of PHYS II.  This implies that the unmatched students were less well prepared in terms of their conceptual understanding of mechanics, but this did not impact their conceptual learning of electricity and magnetism content (as measured by the BEMA).  Additionally, 206 of these students did not appear in any of the rosters for PHYS I in our dataset, meaning they likely did not take PHYS I at CU.  These students scored only slightly lower on the FMCE than students who were unmatched but did appear in one of the PHYS I rosters suggesting that multiple selection effects are at play with respect to which students are matched and unmatched between PHYS I and II.  

While investigation of the reliability of the FMCE was not a goal of this work, our findings also provide a measure of the test-retest reliability of the FMCE.  By measuring student performance at two separate points with a gap large enough to discourage or eliminate students simply remembering their prior responses, our findings also suggest that the FMCE provides measures of students' learning that are reliable over time and multiple implementations.  This is an important measure of test reliability that is often difficult to achieve in practice.  

This work has several important limitations including the fact that data were collected from a single institution.  CU has a relatively high post-instruction FMCE score and a significant fraction of our students score $>90$\% after completing PHYS I.  It is possible that a student population without such high post-instruction scores might show more variation in their responses over time.  Related to this, Thornton argues that students may hold particular views of the world, from the establish physicist view (Newtonian) to the more experienced-based (Aristotelian) view held by most students before any physics instruction.  Our data, combined with the high post-instruction FMCE performance of our students may represent a robust transition in views rather than a retention of knowledge.  In other words, our results may suggest that once a student internalizes a Newtonian view of the world, they do not revert to an Aristotelian one over the time frame of 1-15 months.  If this is the case, students who have not made this transition, but rather have memorized correct patterns of responses over the course of PHYS I, would likely show a much larger drop in performance following a significant passage of time.  

\begin{acknowledgments}
This work was funded by the CU Physics Department.  Special thanks to the faculty and students who participated in the study and to the members of PER@C for all their feedback.  
\end{acknowledgments}

\bibliography{master-refs-12-19}

\begin{thebibliography}{23}%
\makeatletter
\providecommand \@ifxundefined [1]{%
 \@ifx{#1\undefined}
}%
\providecommand \@ifnum [1]{%
 \ifnum #1\expandafter \@firstoftwo
 \else \expandafter \@secondoftwo
 \fi
}%
\providecommand \@ifx [1]{%
 \ifx #1\expandafter \@firstoftwo
 \else \expandafter \@secondoftwo
 \fi
}%
\providecommand \natexlab [1]{#1}%
\providecommand \enquote  [1]{``#1''}%
\providecommand \bibnamefont  [1]{#1}%
\providecommand \bibfnamefont [1]{#1}%
\providecommand \citenamefont [1]{#1}%
\providecommand \href@noop [0]{\@secondoftwo}%
\providecommand \href [0]{\begingroup \@sanitize@url \@href}%
\providecommand \@href[1]{\@@startlink{#1}\@@href}%
\providecommand \@@href[1]{\endgroup#1\@@endlink}%
\providecommand \@sanitize@url [0]{\catcode `\\12\catcode `\$12\catcode
  `\&12\catcode `\#12\catcode `\^12\catcode `\_12\catcode `\%12\relax}%
\providecommand \@@startlink[1]{}%
\providecommand \@@endlink[0]{}%
\providecommand \url  [0]{\begingroup\@sanitize@url \@url }%
\providecommand \@url [1]{\endgroup\@href {#1}{\urlprefix }}%
\providecommand \urlprefix  [0]{URL }%
\providecommand \Eprint [0]{\href }%
\providecommand \doibase [0]{http://dx.doi.org/}%
\providecommand \selectlanguage [0]{\@gobble}%
\providecommand \bibinfo  [0]{\@secondoftwo}%
\providecommand \bibfield  [0]{\@secondoftwo}%
\providecommand \translation [1]{[#1]}%
\providecommand \BibitemOpen [0]{}%
\providecommand \bibitemStop [0]{}%
\providecommand \bibitemNoStop [0]{.\EOS\space}%
\providecommand \EOS [0]{\spacefactor3000\relax}%
\providecommand \BibitemShut  [1]{\csname bibitem#1\endcsname}%
\let\auto@bib@innerbib\@empty
\bibitem [{\citenamefont {Bacon}\ and\ \citenamefont
  {Stewart}(2006)}]{bacon2006retention}%
  \BibitemOpen
  \bibfield  {author} {\bibinfo {author} {\bibfnamefont {Donald~R}\
  \bibnamefont {Bacon}}\ and\ \bibinfo {author} {\bibfnamefont {Kim~A}\
  \bibnamefont {Stewart}},\ }\bibfield  {title} {\enquote {\bibinfo {title}
  {How fast do students forget what they learn in consumer behavior? a
  longitudinal study},}\ }\href@noop {} {\bibfield  {journal} {\bibinfo
  {journal} {Journal of Marketing Education}\ }\textbf {\bibinfo {volume}
  {28}},\ \bibinfo {pages} {181--192} (\bibinfo {year} {2006})}\BibitemShut
  {NoStop}%
\bibitem [{\citenamefont {Rubin}\ and\ \citenamefont
  {Wenzel}(1996)}]{rubin1996forgetting}%
  \BibitemOpen
  \bibfield  {author} {\bibinfo {author} {\bibfnamefont {David~C}\ \bibnamefont
  {Rubin}}\ and\ \bibinfo {author} {\bibfnamefont {Amy~E}\ \bibnamefont
  {Wenzel}},\ }\bibfield  {title} {\enquote {\bibinfo {title} {One hundred
  years of forgetting: A quantitative description of retention.}}\ }\href@noop
  {} {\bibfield  {journal} {\bibinfo  {journal} {Psychological review}\
  }\textbf {\bibinfo {volume} {103}},\ \bibinfo {pages} {734} (\bibinfo {year}
  {1996})}\BibitemShut {NoStop}%
\bibitem [{\citenamefont {Conway}\ \emph {et~al.}(1992)\citenamefont {Conway},
  \citenamefont {Cohen},\ and\ \citenamefont {Stanhope}}]{conway1992retention}%
  \BibitemOpen
  \bibfield  {author} {\bibinfo {author} {\bibfnamefont {Martin~A}\
  \bibnamefont {Conway}}, \bibinfo {author} {\bibfnamefont {Gillian}\
  \bibnamefont {Cohen}}, \ and\ \bibinfo {author} {\bibfnamefont {Nicola}\
  \bibnamefont {Stanhope}},\ }\bibfield  {title} {\enquote {\bibinfo {title}
  {Very long-term memory for knowledge acquired at school and university},}\
  }\href@noop {} {\bibfield  {journal} {\bibinfo  {journal} {Applied cognitive
  psychology}\ }\textbf {\bibinfo {volume} {6}},\ \bibinfo {pages} {467--482}
  (\bibinfo {year} {1992})}\BibitemShut {NoStop}%
\bibitem [{\citenamefont {Freeman}\ \emph {et~al.}(2014)\citenamefont
  {Freeman}, \citenamefont {Eddy}, \citenamefont {McDonough}, \citenamefont
  {Smith}, \citenamefont {Okoroafor}, \citenamefont {Jordt},\ and\
  \citenamefont {Wenderoth}}]{freeman2014active}%
  \BibitemOpen
  \bibfield  {author} {\bibinfo {author} {\bibfnamefont {Scott}\ \bibnamefont
  {Freeman}}, \bibinfo {author} {\bibfnamefont {Sarah~L}\ \bibnamefont {Eddy}},
  \bibinfo {author} {\bibfnamefont {Miles}\ \bibnamefont {McDonough}}, \bibinfo
  {author} {\bibfnamefont {Michelle~K}\ \bibnamefont {Smith}}, \bibinfo
  {author} {\bibfnamefont {Nnadozie}\ \bibnamefont {Okoroafor}}, \bibinfo
  {author} {\bibfnamefont {Hannah}\ \bibnamefont {Jordt}}, \ and\ \bibinfo
  {author} {\bibfnamefont {Mary~Pat}\ \bibnamefont {Wenderoth}},\ }\bibfield
  {title} {\enquote {\bibinfo {title} {Active learning increases student
  performance in science, engineering, and mathematics},}\ }\href@noop {}
  {\bibfield  {journal} {\bibinfo  {journal} {Proceedings of the National
  Academy of Sciences}\ }\textbf {\bibinfo {volume} {111}},\ \bibinfo {pages}
  {8410--8415} (\bibinfo {year} {2014})}\BibitemShut {NoStop}%
\bibitem [{\citenamefont {Francis}\ \emph {et~al.}(1998)\citenamefont
  {Francis}, \citenamefont {Adams},\ and\ \citenamefont
  {Noonan}}]{francis1998retention}%
  \BibitemOpen
  \bibfield  {author} {\bibinfo {author} {\bibfnamefont {Gregory~E}\
  \bibnamefont {Francis}}, \bibinfo {author} {\bibfnamefont {Jeffrey~P}\
  \bibnamefont {Adams}}, \ and\ \bibinfo {author} {\bibfnamefont {Elizabeth~J}\
  \bibnamefont {Noonan}},\ }\bibfield  {title} {\enquote {\bibinfo {title} {Do
  they stay fixed?}}\ }\href@noop {} {\bibfield  {journal} {\bibinfo  {journal}
  {The Physics Teacher}\ }\textbf {\bibinfo {volume} {36}},\ \bibinfo {pages}
  {488--490} (\bibinfo {year} {1998})}\BibitemShut {NoStop}%
\bibitem [{\citenamefont {Bernhard}(2000)}]{bernhard2000ieretention}%
  \BibitemOpen
  \bibfield  {author} {\bibinfo {author} {\bibfnamefont {Jonte}\ \bibnamefont
  {Bernhard}},\ }\bibfield  {title} {\enquote {\bibinfo {title} {Does active
  engagement curricula give long-lived conceptual understanding},}\ }\href@noop
  {} {\bibfield  {journal} {\bibinfo  {journal} {Physics teacher education
  beyond}\ ,\ \bibinfo {pages} {749--752}} (\bibinfo {year}
  {2000})}\BibitemShut {NoStop}%
\bibitem [{\citenamefont {Deslauriers}\ and\ \citenamefont
  {Wieman}(2011)}]{deslauriers2011retention}%
  \BibitemOpen
  \bibfield  {author} {\bibinfo {author} {\bibfnamefont {Louis}\ \bibnamefont
  {Deslauriers}}\ and\ \bibinfo {author} {\bibfnamefont {Carl}\ \bibnamefont
  {Wieman}},\ }\bibfield  {title} {\enquote {\bibinfo {title} {Learning and
  retention of quantum concepts with different teaching methods},}\ }\href
  {\doibase 10.1103/PhysRevSTPER.7.010101} {\bibfield  {journal} {\bibinfo
  {journal} {Phys. Rev. ST Phys. Educ. Res.}\ }\textbf {\bibinfo {volume}
  {7}},\ \bibinfo {pages} {010101} (\bibinfo {year} {2011})}\BibitemShut
  {NoStop}%
\bibitem [{\citenamefont {Finkelstein}\ and\ \citenamefont
  {Pollock}(2005)}]{pollock2005tutorials}%
  \BibitemOpen
  \bibfield  {author} {\bibinfo {author} {\bibfnamefont {N.~D.}\ \bibnamefont
  {Finkelstein}}\ and\ \bibinfo {author} {\bibfnamefont {S.~J.}\ \bibnamefont
  {Pollock}},\ }\bibfield  {title} {\enquote {\bibinfo {title} {Replicating and
  understanding successful innovations: Implementing tutorials in introductory
  physics},}\ }\href {\doibase 10.1103/PhysRevSTPER.1.010101} {\bibfield
  {journal} {\bibinfo  {journal} {Phys. Rev. ST Phys. Educ. Res.}\ }\textbf
  {\bibinfo {volume} {1}},\ \bibinfo {pages} {010101} (\bibinfo {year}
  {2005})}\BibitemShut {NoStop}%
\bibitem [{\citenamefont {Thornton}\ and\ \citenamefont
  {Sokoloff}(1998)}]{thornton1998fmce}%
  \BibitemOpen
  \bibfield  {author} {\bibinfo {author} {\bibfnamefont {Ronald~K}\
  \bibnamefont {Thornton}}\ and\ \bibinfo {author} {\bibfnamefont {David~R}\
  \bibnamefont {Sokoloff}},\ }\bibfield  {title} {\enquote {\bibinfo {title}
  {Assessing student learning of newton’s laws: The force and motion
  conceptual evaluation and the evaluation of active learning laboratory and
  lecture curricula},}\ }\href@noop {} {\bibfield  {journal} {\bibinfo
  {journal} {American Journal of Physics}\ }\textbf {\bibinfo {volume} {66}},\
  \bibinfo {pages} {338--352} (\bibinfo {year} {1998})}\BibitemShut {NoStop}%
\bibitem [{\citenamefont {Wixted}\ and\ \citenamefont
  {Ebbesen}(1991)}]{wixted1991form}%
  \BibitemOpen
  \bibfield  {author} {\bibinfo {author} {\bibfnamefont {John~T}\ \bibnamefont
  {Wixted}}\ and\ \bibinfo {author} {\bibfnamefont {Ebbe~B}\ \bibnamefont
  {Ebbesen}},\ }\bibfield  {title} {\enquote {\bibinfo {title} {On the form of
  forgetting},}\ }\href@noop {} {\bibfield  {journal} {\bibinfo  {journal}
  {Psychological science}\ }\textbf {\bibinfo {volume} {2}},\ \bibinfo {pages}
  {409--415} (\bibinfo {year} {1991})}\BibitemShut {NoStop}%
\bibitem [{\citenamefont {Bjork}(1994)}]{bjork1994memory}%
  \BibitemOpen
  \bibfield  {author} {\bibinfo {author} {\bibfnamefont {Robert~A}\
  \bibnamefont {Bjork}},\ }\href@noop {} {\emph {\bibinfo {title} {Memory and
  metamemory considerations in the training of human beings}}},\ Vol.\ \bibinfo
  {volume} {185}\ (\bibinfo {year} {1994})\BibitemShut {NoStop}%
\bibitem [{\citenamefont {Jackson}\ \emph {et~al.}(2004)\citenamefont
  {Jackson}, \citenamefont {Ventura}, \citenamefont {Chewle},\ and\
  \citenamefont {Graesser}}]{jackson2004tutor}%
  \BibitemOpen
  \bibfield  {author} {\bibinfo {author} {\bibfnamefont {G~Tanner}\
  \bibnamefont {Jackson}}, \bibinfo {author} {\bibfnamefont {Matthew}\
  \bibnamefont {Ventura}}, \bibinfo {author} {\bibfnamefont {Preeti}\
  \bibnamefont {Chewle}}, \ and\ \bibinfo {author} {\bibfnamefont {Art}\
  \bibnamefont {Graesser}},\ }\bibfield  {title} {\enquote {\bibinfo {title}
  {The impact of why/autotutor on learning and retention of conceptual
  physics},}\ }in\ \href@noop {} {\emph {\bibinfo {booktitle} {International
  Conference on Intelligent Tutoring Systems}}}\ (\bibinfo {organization}
  {Springer},\ \bibinfo {year} {2004})\ pp.\ \bibinfo {pages}
  {501--510}\BibitemShut {NoStop}%
\bibitem [{\citenamefont {Johnson-Glenberg}\ \emph {et~al.}(2016)\citenamefont
  {Johnson-Glenberg}, \citenamefont {Megowan-Romanowicz}, \citenamefont
  {Birchfield},\ and\ \citenamefont {Savio-Ramos}}]{johnson2016embodied}%
  \BibitemOpen
  \bibfield  {author} {\bibinfo {author} {\bibfnamefont {Mina~C}\ \bibnamefont
  {Johnson-Glenberg}}, \bibinfo {author} {\bibfnamefont {Colleen}\ \bibnamefont
  {Megowan-Romanowicz}}, \bibinfo {author} {\bibfnamefont {David~A}\
  \bibnamefont {Birchfield}}, \ and\ \bibinfo {author} {\bibfnamefont
  {Caroline}\ \bibnamefont {Savio-Ramos}},\ }\bibfield  {title} {\enquote
  {\bibinfo {title} {Effects of embodied learning and digital platform on the
  retention of physics content: Centripetal force},}\ }\href@noop {} {\bibfield
   {journal} {\bibinfo  {journal} {Frontiers in psychology}\ }\textbf {\bibinfo
  {volume} {7}},\ \bibinfo {pages} {1819} (\bibinfo {year} {2016})}\BibitemShut
  {NoStop}%
\bibitem [{\citenamefont {McDermott}\ and\ \citenamefont
  {Shaffer}(2002)}]{mcdermott2002tutorials}%
  \BibitemOpen
  \bibfield  {author} {\bibinfo {author} {\bibfnamefont {Lillian~C}\
  \bibnamefont {McDermott}}\ and\ \bibinfo {author} {\bibfnamefont {Peter~S}\
  \bibnamefont {Shaffer}},\ }\href@noop {} {\emph {\bibinfo {title} {Tutorials
  in Introductory Physics}}}\ (\bibinfo  {publisher} {Pearson},\ \bibinfo
  {address} {Upper Saddle River, NJ},\ \bibinfo {year} {2002})\BibitemShut
  {NoStop}%
\bibitem [{\citenamefont {Pollock}(2009)}]{pollock2009longitudinal}%
  \BibitemOpen
  \bibfield  {author} {\bibinfo {author} {\bibfnamefont {S.~J.}\ \bibnamefont
  {Pollock}},\ }\bibfield  {title} {\enquote {\bibinfo {title} {Longitudinal
  study of student conceptual understanding in electricity and magnetism},}\
  }\href {\doibase 10.1103/PhysRevSTPER.5.020110} {\bibfield  {journal}
  {\bibinfo  {journal} {Phys. Rev. ST Phys. Educ. Res.}\ }\textbf {\bibinfo
  {volume} {5}},\ \bibinfo {pages} {020110} (\bibinfo {year}
  {2009})}\BibitemShut {NoStop}%
\bibitem [{\citenamefont {Shaffer}\ and\ \citenamefont
  {McDermott}(1992)}]{shaffer1992circuits}%
  \BibitemOpen
  \bibfield  {author} {\bibinfo {author} {\bibfnamefont {Peter~S}\ \bibnamefont
  {Shaffer}}\ and\ \bibinfo {author} {\bibfnamefont {Lillian~C}\ \bibnamefont
  {McDermott}},\ }\bibfield  {title} {\enquote {\bibinfo {title} {Research as a
  guide for curriculum development: An example from introductory electricity.
  part ii: Design of instructional strategies},}\ }\href@noop {} {\bibfield
  {journal} {\bibinfo  {journal} {American Journal of Physics}\ }\textbf
  {\bibinfo {volume} {60}},\ \bibinfo {pages} {1003--1013} (\bibinfo {year}
  {1992})}\BibitemShut {NoStop}%
\bibitem [{\citenamefont {McDermott}\ \emph {et~al.}(2000)\citenamefont
  {McDermott}, \citenamefont {Shaffer},\ and\ \citenamefont
  {Constantinou}}]{mcdermott2000circuits}%
  \BibitemOpen
  \bibfield  {author} {\bibinfo {author} {\bibfnamefont {Lillian~C}\
  \bibnamefont {McDermott}}, \bibinfo {author} {\bibfnamefont {Peter~S}\
  \bibnamefont {Shaffer}}, \ and\ \bibinfo {author} {\bibfnamefont
  {Constantinos~P}\ \bibnamefont {Constantinou}},\ }\bibfield  {title}
  {\enquote {\bibinfo {title} {Preparing teachers to teach physics and physical
  science by inquiry},}\ }\href@noop {} {\bibfield  {journal} {\bibinfo
  {journal} {Physics Education}\ }\textbf {\bibinfo {volume} {35}},\ \bibinfo
  {pages} {411} (\bibinfo {year} {2000})}\BibitemShut {NoStop}%
\bibitem [{\citenamefont {Crouch}\ and\ \citenamefont
  {Mazur}(2001)}]{crouch2001pi}%
  \BibitemOpen
  \bibfield  {author} {\bibinfo {author} {\bibfnamefont {Catherine~H}\
  \bibnamefont {Crouch}}\ and\ \bibinfo {author} {\bibfnamefont {Eric}\
  \bibnamefont {Mazur}},\ }\bibfield  {title} {\enquote {\bibinfo {title} {Peer
  instruction: Ten years of experience and results},}\ }\href@noop {}
  {\bibfield  {journal} {\bibinfo  {journal} {American journal of physics}\
  }\textbf {\bibinfo {volume} {69}},\ \bibinfo {pages} {970--977} (\bibinfo
  {year} {2001})}\BibitemShut {NoStop}%
\bibitem [{\citenamefont {https://www.pearsonmylabandmastering.com/northame
  rica/masteringphysics/}(2020)}]{MasteringPhysicswebsite}%
  \BibitemOpen
  \bibfield  {author} {\bibinfo {author} {\bibnamefont
  {https://www.pearsonmylabandmastering.com/northame rica/masteringphysics/}},\
  }\href@noop {} {} (\bibinfo {year} {2020})\BibitemShut {NoStop}%
\bibitem [{\citenamefont
  {{https://www.flipitphysics.com/}}(2020)}]{flipitwebsite}%
  \BibitemOpen
  \bibfield  {author} {\bibinfo {author} {\bibnamefont
  {{https://www.flipitphysics.com/}}},\ }\href@noop {} {} (\bibinfo {year}
  {2020})\BibitemShut {NoStop}%
\bibitem [{\citenamefont {Von~Korff}\ \emph {et~al.}(2016)\citenamefont
  {Von~Korff}, \citenamefont {Archibeque}, \citenamefont {Gomez}, \citenamefont
  {Heckendorf}, \citenamefont {McKagan}, \citenamefont {Sayre}, \citenamefont
  {Schenk}, \citenamefont {Shepherd},\ and\ \citenamefont
  {Sorell}}]{vonkorff2016pedagogy}%
  \BibitemOpen
  \bibfield  {author} {\bibinfo {author} {\bibfnamefont {Joshua}\ \bibnamefont
  {Von~Korff}}, \bibinfo {author} {\bibfnamefont {Benjamin}\ \bibnamefont
  {Archibeque}}, \bibinfo {author} {\bibfnamefont {K~Alison}\ \bibnamefont
  {Gomez}}, \bibinfo {author} {\bibfnamefont {Tyrel}\ \bibnamefont
  {Heckendorf}}, \bibinfo {author} {\bibfnamefont {Sarah~B}\ \bibnamefont
  {McKagan}}, \bibinfo {author} {\bibfnamefont {Eleanor~C}\ \bibnamefont
  {Sayre}}, \bibinfo {author} {\bibfnamefont {Edward~W}\ \bibnamefont
  {Schenk}}, \bibinfo {author} {\bibfnamefont {Chase}\ \bibnamefont
  {Shepherd}}, \ and\ \bibinfo {author} {\bibfnamefont {Lane}\ \bibnamefont
  {Sorell}},\ }\bibfield  {title} {\enquote {\bibinfo {title} {Secondary
  analysis of teaching methods in introductory physics: A 50 k-student
  study},}\ }\href@noop {} {\bibfield  {journal} {\bibinfo  {journal} {American
  Journal of physics}\ }\textbf {\bibinfo {volume} {84}},\ \bibinfo {pages}
  {969--974} (\bibinfo {year} {2016})}\BibitemShut {NoStop}%
\bibitem [{\citenamefont {Ding}\ \emph {et~al.}(2006)\citenamefont {Ding},
  \citenamefont {Chabay}, \citenamefont {Sherwood},\ and\ \citenamefont
  {Beichner}}]{ding2006bema}%
  \BibitemOpen
  \bibfield  {author} {\bibinfo {author} {\bibfnamefont {Lin}\ \bibnamefont
  {Ding}}, \bibinfo {author} {\bibfnamefont {Ruth}\ \bibnamefont {Chabay}},
  \bibinfo {author} {\bibfnamefont {Bruce}\ \bibnamefont {Sherwood}}, \ and\
  \bibinfo {author} {\bibfnamefont {Robert}\ \bibnamefont {Beichner}},\
  }\bibfield  {title} {\enquote {\bibinfo {title} {Evaluating an electricity
  and magnetism assessment tool: Brief electricity and magnetism assessment},}\
  }\href {\doibase 10.1103/PhysRevSTPER.2.010105} {\bibfield  {journal}
  {\bibinfo  {journal} {Phys. Rev. ST Phys. Educ. Res.}\ }\textbf {\bibinfo
  {volume} {2}},\ \bibinfo {pages} {010105} (\bibinfo {year}
  {2006})}\BibitemShut {NoStop}%
\bibitem [{\citenamefont {Thornton}\ \emph {et~al.}(2009)\citenamefont
  {Thornton}, \citenamefont {Kuhl}, \citenamefont {Cummings},\ and\
  \citenamefont {Marx}}]{thornton2009fmce}%
  \BibitemOpen
  \bibfield  {author} {\bibinfo {author} {\bibfnamefont {Ronald~K}\
  \bibnamefont {Thornton}}, \bibinfo {author} {\bibfnamefont {Dennis}\
  \bibnamefont {Kuhl}}, \bibinfo {author} {\bibfnamefont {Karen}\ \bibnamefont
  {Cummings}}, \ and\ \bibinfo {author} {\bibfnamefont {Jeffrey}\ \bibnamefont
  {Marx}},\ }\bibfield  {title} {\enquote {\bibinfo {title} {Comparing the
  force and motion conceptual evaluation and the force concept inventory},}\
  }\href {\doibase 10.1103/PhysRevSTPER.5.010105} {\bibfield  {journal}
  {\bibinfo  {journal} {Phys. Rev. ST Phys. Educ. Res.}\ }\textbf {\bibinfo
  {volume} {5}},\ \bibinfo {pages} {010105} (\bibinfo {year}
  {2009})}\BibitemShut {NoStop}%
\end{thebibliography}%

\end{document}